
\magnification=1200
\baselineskip=13pt
\tolerance=100000
\overfullrule=0pt
\rightline{UR-1325\ \ \ \ \ \ \ }
\rightline{ER-40685-775}

\baselineskip=20pt
\centerline{\bf ON THE HIGHER ORDER CORRECTIONS TO}
\centerline{\bf HEAVY QUARK EFFECTIVE THEORY}

\bigskip
\medskip

\centerline{Ashok Das}
\centerline{Department of Physics and Astronomy}
\centerline{University of Rochester}
\centerline{Rochester, NY 14627}
\bigskip
\bigskip
\bigskip
\bigskip
\centerline{\underbar{Abstract}}
\medskip

We show that the different ways of deriving the Heavy Quark Effective Theory
(HQET)  lead to equivalent theories.  The equivalence can be established
through a careful redefinition of the field variables.  We demonstrate the
equivalence to order ${1 \over m^5}$ in the presence of a constant electric
field.

\vfill\eject

\noindent {\bf I. Introduction:}

\medskip

In recent years, there has been a lot of work in the context of Heavy Quark
Effective Theories (HQET) [1-6].
  When a quark is heavy, it can be effectively
described by a two component spinor and there are essentially two different
ways of obtaining an effective theory for such a quark.  The most logical
way to take the heavy quark mass limit, in some sense, is through the
Foldy-Wouthuysen transformations [7,8] where one essentially diagonalizes the
quark Lagrangian such that the upper and lower two component spinors
decouple after which one restricts to the positive energy two component
upper spinors.  (In the functional language, one would simply integrate out
the lower component spinors after diagonalization and absorb the result
into the normalization factor.)  The other approach to an effective theory
is the traditional way [4,9] where one decomposes
 the four component spinor into
a ``large" and a ``small" two component spinor.  One then eliminates
the ``small"
component through its equation of motion to obtain an effective theory for
the ``large" component spinors.  (In the functional language, one would
integrate out the ``small" components and absorb the result into the
normalization factor of the path integral.) While the two methods give the
same identical theory in the lowest order in ${1 \over m}$, their
structures appear to be different at higher orders.  This disagreement has
justifiably raised concern in the recent literature [10] mainly because higher
order corrections to various physical processes are currently being
calculated [11] using the theory
 resulting from the traditional method of eliminating
the ``small" components.

The discrepancy between the resulting theories in the two approaches is
nothing new.  It was already noted in connection with an electron
interacting with an external electric field [12] as well as in the context of
the Tamm-Dancoff method in nuclear physics [13], that eliminating
the ``small"
component naively leads to a nonhermitian Hamiltonian.  (In the case of the
electron interacting with an external electric field, the
 lowest order manifestation of the nonhermiticity is
in an imaginary electric dipole moment.)  To the best of our knowledge, the
resolution of this puzzle was first proposed in
 ref.~14 where it was noted that
a renormalization (redefinition) of the large components is essential for
the hermiticity of the Hamiltonian.  In the case of the Dirac electron
 interacting with an electric field, the
equivalence of the two approaches was thereby demonstrated to order
${1 \over m^2}$ [14].

It is in general believed that an appropriate field redefinition will lead
to an equivalence of the two approaches.  In this note, we wish to
demonstrate up to order ${1 \over m^5}$ that both the methods indeed give
the same theory with appropriate field redefinitions.  Both the methods
have their advantages and disadvantages and we comment on this in the
conclusion.  The equivalence of the two approaches is quite important and
so is the understanding of
 the field redefinition since otherwise the higher order
corrections calculated with the
 naive effective theory may not represent the true
physical effects.  In sec. II, we review the known results up to order
${1 \over m^2}$ and try to bring out the necessity for a field
redefinition.  In sec. III, we show that, in the case of
the free theory,
the two approaches lead to the same effective theory upon field
redefinition.  We then demonstrate to order ${1 \over m^5}$ that in the
presence of a constant electric field, the two approaches also give the
same effective theory through a careful redefinition of fields.  In sec.
IV, we show the equivalence of the two approaches in
the functional approach and present our conclusions in sec. V.
\medskip

\noindent {\bf II. The Problem of Imaginary Dipole Moment:}

\medskip

In this section, we consider a Dirac electron interacting with a constant,
external electric field.  While the discussion can be carried out equally
well in the first or second quantized language we will follow a first
quantized approach for simplicity and clarity.  Our discussion in this
section will follow closely the work in ref.~14.  The Dirac equation in the
present case has the form (we use Bjorken-Drell metric,
$\gamma^0 = \pmatrix{I &0\cr
0 &I\cr} ,\ \vec \gamma = \pmatrix{0 &\vec \sigma\cr
-\vec \sigma &0\cr}$)
$$\left( i \gamma^0 \left( \partial_0 + i e A_0 \right) + i \vec
\gamma \cdot \vec \nabla - m \right) \psi = 0 \eqno(2.1)$$
with
$$\vec E = \vec \nabla A_0 = \ {\rm constant} \eqno(2.2)$$
If we let
$$\psi \rightarrow e^{-imt} \psi \eqno(2.3)$$
then Eq. (2.1) takes the form
$$\left( i \gamma^0 \left( \partial_0 + ie A_0 \right) + i
\vec \gamma \cdot \vec \nabla
 - m \left( 1 - \gamma^0 \right) \right) \psi = 0 \eqno(2.4)$$
Introducing the ``large" and the ``small" components as
 (In the current terminology of the subject our entire discussion will be
with the choice $v^\mu = (1,0,0,0)$ for simplicity.)
$$\psi = \pmatrix{\psi_L\cr
\noalign{\vskip 5pt}%
\psi_S\cr}\eqno(2.5)$$
we note that the Dirac equation (2.4) separates into two equations
$$\psi_S = \left( i \partial_0 - e A_0 + 2m \right)^{-1}
\left( - i \vec \sigma \cdot \vec \nabla \right) \psi_L =
A \psi_L \eqno(2.6)$$
$$\left( i \partial_0 - e A_0 \right) \psi_L = - i \vec \sigma
\cdot \vec \nabla \psi_S \eqno(2.7)$$
Upon substituting Eq. (2.6) into Eq. (2.7) and expanding up to order
${1 \over m^2}$, the equation for the ``large" component takes the form
$$\left( i \partial_0 - e A_0 \right)\psi_L =
\left( - {1 \over 2m} \ \vec \nabla^2 -
 {ie \over 4m^2} \ \vec \sigma \cdot
\left( \vec E \times \vec \nabla \right)
+ {e \over 4m^2} \ \vec E \cdot \vec \nabla  + O \left( {1 \over m^3} \right)
 \right)
\psi_L \eqno(2.8)$$

It is the last term on the right hand side in Eq. (2.8) which represents an
imaginary electric dipole moment and arises naturally as a consequence of
eliminating the ``small" component spinors.  While this may be puzzling,
it's origin is not hard to understand.  As explained beautifully in ref.~14,
one can view the process of eliminating the ``small" component also
equivalently as finding a transformation which will take
$$\psi = \pmatrix{\psi_L\cr
\noalign{\vskip 5pt}%
\psi_S\cr} = \pmatrix{\psi_L \cr
\noalign{\vskip 5pt}%
A\psi_L\cr} \buildrel S \over \longrightarrow  \pmatrix{\psi_L\cr
\noalign{\vskip 5pt}%
0\cr} \eqno(2.9)$$
Such a transformation is generated by the matrix
$$S = \pmatrix{I &0\cr
\noalign{\vskip 5pt}%
-A &I\cr} \eqno(2.9^\prime)$$

It is clear that this transformation is not unitary and as a
result the Hamiltonian does not remain hermitian under such a
transformation.  Another manifestation of
 the transformation  not being unitary is to note that under such a
transformation, the norm of the state
 is not invariant.  In fact, let us note that
$$\eqalign{\int d^3x \ \psi^\dagger \psi &= \int d^3x
\big( \psi^\dagger_L \psi_L + \psi^\dagger_S \psi_S \big)\cr
&= \int d^3x \ \psi^\dagger_L \big( 1 + A^\dagger A \big) \psi_L
\not= \int d^3x\ \psi^\dagger_L \psi_L\cr}
\eqno(2.10)$$
Therefore, if
$$\int d^3x\ \psi^\dagger \psi = 1 \eqno(2.11)$$
then,
$$\int d^3x \ \psi^\dagger_L \psi_L \not= 1 \eqno(2.12)$$
It is clear, however, that the norm can be maintained (state will be
normalized) if we redefine
$$\widetilde \psi_L = \left( 1 + A^\dagger A \right)^{1/2} \psi_L
\eqno(2.13)$$
As is described in ref.~14, this redefinition also restores hermiticity of
the Hamiltonian.
 (The norm will, of course, be time independent.)
  In fact, let us note from Eqs. (2.6) and (2.13) that if
we define up to order ${1 \over m^2}$
$$\eqalign{\widetilde \psi_L &= \bigg( 1 - {1 \over 4m^2}\ \vec \nabla^2
\bigg)^{1/2} \psi_L \simeq \bigg( 1 - {1 \over 8m^2}\ \vec \nabla^2
\bigg) \psi_L\cr
{\rm or,}\qquad\qquad \psi_L &\simeq \bigg( 1 + {1 \over 8m^2}\ \vec
\nabla^2 \bigg) \widetilde \psi_L \cr}\eqno(2.14)$$
then Eq. (2.8) would take the form
$$\left( i \partial_0 - e A_0 \right) \widetilde \psi_L =
\left( - {1 \over 2m}\ \vec \nabla^2 - {ie \over 4m^2}\
\vec \sigma \cdot \left( \vec E \times \vec \nabla \right)
 + O \left( {1 \over m^3}\right) \right)
\widetilde \psi_L
\eqno(2.15)$$
The imaginary electric dipole term has now been absorbed into the field
redefinition and as we will see in the next section this is exactly the
same equation which we would get through the Foldy-Wouthuysen
transformation up to order ${1 \over m^2}$.

\medskip

\noindent {\bf III. Equivalence up to Order ${1 \over m^5}$:}

\medskip

As a warm up, let us consider a free Dirac particle satisfying
$$\left( i \gamma^0 \partial_0 + i \vec \gamma \cdot
\vec \nabla - m \right) \psi = 0 \eqno(3.1)$$
The idea behind the Foldy-Wouthuysen transformations [7] is to block
diagonalize the Dirac operator such that the upper and the lower two
component spinors are decoupled.  It is well known [7] that under
 the unitary transformation
$$\psi \rightarrow \psi^\prime = U \psi
 = e^{{i \over 2m}\ \vec \gamma
 \cdot \vec \nabla \alpha ( {| \vec \nabla | \over m} )}
\ \psi \eqno(3.2)$$
where
$$\alpha \left( {| \vec \nabla | \over m} \right) = {m \over | \vec
\nabla |} \ \tanh^{-1} \left( {| \vec \nabla | \over m} \right)
\eqno(3.3)$$
the Dirac operator transforms into the diagonal form
$$\eqalign{\big( i \gamma^0 \partial_0 + i \vec \gamma
\cdot \vec \nabla - m \big) &\rightarrow U \big( i \gamma^0
\partial_0 + i \vec \gamma \cdot \vec \nabla - m
\big) U^\dagger\cr
&= i \gamma^0 \partial_0 - \big( m^2 - \vec
\nabla^2 \big)^{1/2} \cr}\eqno(3.4)$$
Thus the transformed equation has the form
$$\left( i \gamma^0 \partial_0 - \left( m^2 - \vec \nabla^2 \right)^{1/2}
\right) \psi^\prime = 0 \eqno(3.5)$$
If we further let
$$\psi^\prime \rightarrow e^{-imt} \psi^\prime \eqno(3.6)$$
then Eq. (3.5) will take the form
$$\left( i \gamma^0 \partial_0 + m \gamma^0 -
\left( m^2 - \vec \nabla^2 \right)^{1/2} \right)
\psi^\prime = 0 \eqno(3.7)$$
Writing
$$\psi^\prime = \pmatrix{\psi_1\cr
\noalign{\vskip 5pt}%
\psi_2\cr} \eqno(3.8)$$
and restricting to the upper two component spinor, we obtain from
Eq. (3.7)
$$\left( i \partial_0 + m - \left( m^2 - \vec \nabla^2 \right)^{1/2}
\right) \psi_1 = 0 \eqno(3.9)$$
A power series expansion of Eq. (3.9) in ${1 \over m}$ would then give the
effective dynamical equation in the Foldy-Wouthuysen approach.

In contrast, let us note from Eqs. (2.6) and (2.7) that in the absence of
interactions
$$\eqalignno{\psi_S &= A \psi_L = (i \partial_0 + 2m )^{-1}
 (-i \vec \sigma \cdot \vec \nabla ) \psi_L &(3.10)\cr
i \partial_0 \psi_L &= -i \vec \sigma \cdot \vec \nabla \psi_S =
- \vec \sigma \cdot \vec \nabla (i \partial_0 + 2m )^{-1}
\vec \sigma \cdot \vec \nabla \psi_L= - (i \partial_0 + 2m )^{-1}
\vec \nabla^2 \psi_L &(3.11)\cr}$$
The equation for the ``large" component, Eq. (3.11), of course, does not at
all resemble Eq. (3.9) obtained through the other approach.  Let us also
note that if we define
$$\widetilde \psi_L = \left( 1 + A^\dagger A \right)^{1/2}
\psi_L = \left( 1 - \left( i \partial_0 + 2m \right)^{-2}
\vec \nabla^2 \right)^{1/2} \psi_L \eqno(3.12)$$
as we should to maintain normalization, the form of the equation does not
change
$$i \partial_0 \widetilde \psi_L = - \left( i \partial_0
+ 2m \right)^{-1} \vec \nabla^2 \widetilde \psi_L \eqno(3.13)$$
This is mainly because the redefinition factor in Eq. (3.12) is field
independent, commutes with $\partial_0$ as well as $\vec \nabla$ and that
Eq. (3.11) is linear in $\psi_L$.  However, we note that Eq. (3.13) can
also be written as
$$\big( i \partial_0 + ( i \partial_0 + 2m )^{-1} \vec \nabla^2
\big) \widetilde \psi_L = 0 \eqno(3.14)$$
Therefore, if we further redefine
$$\widetilde \psi_L \rightarrow
e^{-i(m-(m^2- \vec \nabla^2)^{1/2} -
 ( i \partial_0 + 2m )^{-1}
\vec \nabla^2) t}
 \widetilde \psi_L \eqno(3.15)$$
then Eq. (3.14) will take the form
$$\left( i \partial_0 + m - \left( m^2 -
\vec \nabla^2 \right)^{1/2} \right) \widetilde \psi_L = 0 \eqno(3.16)$$
which is, of course, what we have in the Foldy-Wouthuysen approach
 (see Eq. (3.9)).
 The point of this exercise
is to note that even after the spinor is normalized properly, there still
remains an arbitrariness up to unitary transformations [15] and this is crucial
in establishing the equivalence of the two approaches.  Alternately, we
note that Eq. (3.14) can also be written as
$$\left( i \partial_0 + 2m \right)^{-1}
\left( i \partial_0 + m + \left( m^2 - \vec \nabla^2
\right)^{1/2} \right) \left( i \partial_0 + m - \left(
m^2 - \vec \nabla^2 \right)^{1/2} \right) \widetilde \psi_L= 0
\eqno(3.16^\prime)$$
Consequently, if we redefine
$$\widetilde \psi_L \rightarrow \left( i \partial_0 + 2m \right)^{-1/2}
\left( i \partial_0 + m + \left( m^2 - \vec \nabla^2 \right)^{1/2}
\right)^{1/2} \widetilde \psi_L
\eqno(3.16^{\prime \prime})$$
then the dynamical equation would take the
form
$$\left( i \partial_0 + m - \left( m^2 - \vec \nabla^2 \right)^{1/2}
\right) \widetilde \psi_L = 0$$
which is the equation obtained through the Foldy-Wouthuysen approach.  It
would seem, therefore, that at the level of the equation of motion, there is an
arbitrariness in the choice of field redefinition.  Furthermore, the
redefinition in Eq. (3.16$^{\prime \prime}$) would seem to imply a change
in the norm of the state.  However, in the next section, where we will show
the equivalence of the two approaches for the free theory in the functional
approach, it will become clear that such a field redefinition is essential
for the matching of the functional determinants.

Let us next consider a Dirac particle interacting with a constant, external
electric field.  This is chosen to avoid the
 unnecessary complexities associated with a
generalized non Abelian gauge potential which do not provide any new
insight into the problem.  The equation of motion is
$$\left( i \gamma^0 \left( \partial_0 + ie A_0 \right) +
i \vec \gamma \cdot \vec \nabla - m \right) \psi = 0 \eqno(3.17)$$
with
$$\vec E = \vec \nabla A_0 = \ {\rm constant}\eqno(3.18)$$
In the presence of interactions, the Foldy-Wouthuysen transformations have
to be implemented order by order.  We refer the readers to
the literature [16]  for
details
and merely quote here the result that the equation satisfied by the upper
two component spinor in the present case, up to order ${1 \over
 m^5}$, after a phase redefinition of the form in Eq. (3.6) is given by
$$\eqalign{\big( i \partial_0 - e A_0 \big) \psi_1 = \bigg(
&- {1 \over 2m}\ \vec \nabla^2 - {ie \over 4m^2}\ \vec \sigma \cdot
\big( \vec E \times \vec \nabla \big) -
{1 \over 8m^3}\ \big( \vec \nabla^2 \big)^2 +
{e^2 \over 8m^3}\ \vec E^2\cr
&- {3ie \over 16m^4}\ \vec \sigma \cdot \big( \vec E \times \vec \nabla
\big)\  \vec \nabla^2 - {1 \over 16m^5}\ \big( \vec \nabla^2 \big)^3\cr
&+ {5 e^2 \over 48 m^5}\ \vec E^2 \vec \nabla^2 - {e^2 \over
24m^5}\ \big( \vec E \cdot \vec \nabla \big)^2  +
O \big( {1 \over m^6}\big)  \bigg)
 \psi_1 \cr}\eqno(3.19)$$
Note here that up to order ${1 \over m^2}$, it is the same equation as in
Eq. (2.15).

To analyze the effective equation in the other approach, let us recall from
Eqs. (2.6) and (2.7) that in the present case
$$\psi_S = A \psi_L = - \left( i \partial_0 - e A_0
+ 2m \right)^{-1} \vec \sigma
\cdot \vec \nabla \psi_L$$
$$\left( i \partial_0 - eA_0 \right) \psi_L = - i \vec \sigma \cdot
\vec \nabla \psi_S = - \vec \sigma \cdot \vec \nabla
\left( i \partial_0 - eA_0 + 2m \right)^{-1}
\vec \sigma \cdot \vec \nabla \psi_L$$
The normalization factor, in this case,
$$\left( 1+ A^\dagger A \right)^{1/2} = \left( 1 - \vec \sigma \cdot
\vec \nabla \left( i \partial_0 - eA_0
+ 2m \right)^{-2} \vec \sigma \cdot \vec \nabla \right)^{1/2} \eqno(3.20)$$
is field dependent and involves the Dirac operator.  Expanding to order
${1 \over m^5}$, we note that the redefined field will be given by
$$\eqalign{\widetilde \psi_L &= \big( 1 + A^\dagger A \big)^{1/2}
\psi_L\cr
{\rm or,}\qquad \psi_L &= \big( 1 + A^\dagger A \big)^{-1/2}
\widetilde \psi_L\cr
&= \bigg[ 1 + {1 \over 8m^2}\ \vec \nabla^2 -
{1 \over 8m^3}\ \vec \sigma \cdot \vec \nabla \big( i \partial_0
- eA_0 \big) \vec \sigma \cdot \vec \nabla +
{3 \over 128 m^4}\ \big( \vec \nabla^2 \big)^2\cr
&\qquad + {3 \over 32m^4}\ \vec \sigma \cdot \vec \nabla
\big( i \partial_0
- eA_0 \big)^2 \vec \sigma \cdot \vec \nabla -
{1 \over 16 m^5}\ \vec \sigma \cdot
 \vec \nabla \big( i \partial_0 - eA_0 \big)^3
\vec \sigma \cdot \vec \nabla\cr
&\qquad - {3 \over 128 m^5}\ \vec \nabla^2\  \vec \sigma \cdot
\vec \nabla \big( i \partial_0 - eA_0 \big) \vec \sigma \cdot \vec \nabla
\cr
&\qquad  - {3 \over 128 m^5}\ \vec \sigma \cdot \vec \nabla
\big( i \partial_0 - e A_0 \big)
\vec \sigma \cdot \vec \nabla \  \vec \nabla^2
+ O \bigg( {1 \over m^6}\bigg) \bigg] \widetilde \psi_L\cr}\eqno(3.21)$$
Substituting this back into the equation for the ``large" component we
obtain after some algebra
$$\eqalign{\big( i \partial_0 - e A_0 \big) \widetilde \psi_L = \bigg[
&- {1 \over 2m} \ \vec \nabla^2 - {ie \over 4m^2}\
\vec \sigma \cdot \big( \vec E \times \vec \nabla \big) -
{1 \over 16m^3}\ \big( \vec \nabla^2\big)^2 + {e^2 \over
8m^3}\ \vec E^2\cr
&+ {e \over 32 m^4}\ \big( 2 \vec E \cdot \vec \nabla -
3 i \vec \sigma \cdot \big( \vec E \times \vec \nabla \big)\big)
\vec \nabla^2 - {3 \over 256 m^5}\
\big( \vec \nabla^2 \big)^3\cr
&+ {e^2 \over 16 m^5}\  \vec E^2  \vec \nabla^2 +
{e^2 \over 16m^5}\
\big( \vec E \cdot \vec \nabla \big)^2 +
{ie^2  \over 32 m^5}\
\vec E \cdot \vec \nabla \ \vec \sigma \cdot
\big( \vec E \times \vec \nabla \big)\cr
&+ {1 \over 8 m^2}\ \vec \nabla^2
\big( i \partial_0 - eA_0 \big) + {e \over 8m^3}\
 \big( \vec E \cdot \vec \nabla + i \vec \sigma \cdot \big( \vec E
\times \vec \nabla \big)\big) \big(i \partial_0
-eA_0 \big)\cr
&+ {9 \over 128 m^4}\ \big( \vec \nabla^2 \big)^2 \big( i \partial_0 - eA_0
\big) - {3e^2 \over 16m^4}\ \vec E^2 \big( i \partial_0 - eA_0 \big)\cr
&- {3e \over 16m^4}\ \vec E \cdot \vec \nabla \big( i \partial_0 -eA_0 \big
)^2 - {1 \over 32m^4}\ \vec \nabla^2 \big( i \partial_0 - eA_0 \big)^3\cr
&+ \dots\dots \bigg] \widetilde \psi_L\cr}\eqno(3.22)$$

We note here that since the normalization factor involves the Dirac
operator, Eq. (3.22) has to be solved iteratively.  This is crucial because
this implies that one must keep terms consistently up to any given order.
 We note that this is a new feature not present at order ${1 \over m^2}$.
Iterating Eq. (3.22) up to order ${1 \over m^5}$ gives
$$\eqalign{\big( i \partial_0 - e A_0 \big) \widetilde \psi_L = \bigg[
&- {1 \over 2m} \ \vec \nabla^2 - {ie \over 4m^2}\
\vec \sigma \cdot \big( \vec E \times \vec \nabla \big) -
{1 \over 8m^3}\ \big( \vec \nabla^2\big)^2 + {e^2 \over
8m^3}\ \vec E^2\cr
&- {3ie \over 16 m^4}\
\vec \sigma \cdot \big( \vec E \times \vec \nabla \big)
\vec \nabla^2 - {1 \over 16 m^5}\
\big( \vec \nabla^2 \big)^3 + {15e^2 \over 64 m^5}\
\vec E^2 \vec \nabla^2\cr
&+ {7e^2 \over 32m^5}\ \big(\vec E \cdot \vec \nabla\big)^2 + O
\bigg({1 \over m^6}\bigg) \bigg] \widetilde \psi_L \cr}\eqno(3.23)$$
Comparing with Eq. (3.19), we see that the two equations almost agree -- in
fact, only the coefficients of the last two terms are different.  However,
as discussed earlier,
we also recognize that $\widetilde \psi_L$ is unique only up to a unitary
transformation.  Taking advantage of this, we let
$$\widetilde \psi_L \rightarrow e^{{25 e \over 192 m^5}\
\vec E \cdot \vec \nabla \vec \nabla^2} \widetilde \psi_L \eqno(3.24)$$
With this redefinition, we note that Eq. (3.23) takes the form
$$\eqalign{\big( i \partial_0 - e A_0 \big) \widetilde \psi_L = \bigg[
&- {1 \over 2m} \ \vec \nabla^2 - {ie \over 4m^2}\
\vec \sigma \cdot \big( \vec E \times \vec \nabla \big) -
{1 \over 8m^3}\ \big( \vec \nabla^2\big)^2 + {e^2 \over
8m^3}\ \vec E^2\cr
&- {3ie \over 16 m^4}\
\vec \sigma \cdot \big( \vec E \times \vec \nabla \big)
\vec \nabla^2 - {1 \over 16 m^5}\
\big( \vec \nabla^2 \big)^3 + {5e^2 \over 48 m^5}\
\vec E^2 \vec \nabla^2\cr
&- {e^2 \over 24m^5}\ \big(\vec E \cdot \vec \nabla\big)^2 + O
\bigg({1 \over m^6}\bigg)
\bigg] \widetilde \psi_L \cr}\eqno(3.25)$$
This is exactly the same as Eq. (3.19) which is what we would obtain in the
Foldy-Wouthuysen approach.  Thus, the two approaches give the same
effective equation up to order ${1 \over m^5}$ with appropriate field
redefinition.

\medskip

\noindent {\bf IV. Functional Equivalence:}

\medskip

In this section, we will show the equivalence of the two approaches in the
functional method.  Let us start with a free Dirac theory described by
$${\cal L} = \overline \psi \left( i \rlap\slash{\partial} - m
\right) \psi \eqno(4.1)$$
If we make the phase transformation
$$\psi \rightarrow e^{-imt} \psi \eqno(4.1^\prime)$$
then the Lagrangian will take the form
$${\cal L} = \overline \psi \left( i \rlap\slash{\partial} - m
\left( 1 - \gamma^0 \right) \right) \psi \eqno(4.2)$$
The generating functional in this case is given by
$$Z = N \int {\cal D} \overline \psi
{\cal D} \psi\  e^{i \int d^4x\ {\cal L}} \eqno(4.2)$$
In the Foldy-Wouthuysen approach if we make a unitary redefinition of the
field variables
$$\psi^\prime = e^{{i \over 2m}\ \vec \gamma \cdot \vec \nabla \alpha
({|\vec \nabla | \over m})} \psi =
\pmatrix{\psi_1\cr
\noalign{\vskip 5pt}
\psi_2\cr} \eqno(4.3)$$
with
$$\alpha \left( {| \vec \nabla | \over m}\right) = {m \over |
\vec \nabla |} \tanh^{-1}
\left( {| \vec \nabla | \over m}\right) \eqno(4.4)$$
then the Lagrangian diagonalizes and takes the form
$$\eqalign{{\cal L} = \psi^\dagger_1 \big(i \partial_0 +m
 - \big( m^2 -
\vec \nabla^2\big)^{1/2} \big)&\psi_1 +
\psi^\dagger_2 \big( i \partial_0 + m
 + \big( m^2 - \vec \nabla^2 \big)^{1/2}
\big) \psi_2\cr
&+ \ {\rm surface\ terms}\cr}\eqno(4.5)$$
The functional measure does not change under the unitary redefinition of
the fields in Eq. (4.3).  Consequently, the generating functional has the
form
$$\eqalign{Z &= N \int {\cal D} \psi^\dagger_1 {\cal D} \psi_1 {\cal D}
\psi^\dagger_2 {\cal D} \psi_2
e^{i \int d^4x \big( \psi^\dagger_1 (i \partial_0 + m -
(m^2 - \vec \nabla^2)^{1/2}) \psi_1 + \psi^\dagger_2
(i \partial_0 + m + (m^2 - \vec \nabla^2 )^{1/2}) \psi_2\big)}\cr
&= N \big[ \det
\big( i \partial_0 + m
 + (m^2 - \vec \nabla^2 )^{1/2}\big)\big] \int
 {\cal D} \psi^\dagger_1 {\cal D} \psi_1
e^{i \int d^4x \big( \psi^\dagger_1 (i \partial_0 + m -
(m^2 - \vec \nabla^2)^{1/2}) \psi_1 \big)}\cr
&= N \big[ \det
\big( i \partial_0 + m + (m^2 - \vec \nabla^2 )^{1/2}\big)\big]
\ Z_{FW}\cr}\eqno(4.6)$$
We note here that the propagator in the Foldy-Wouthuysen approach would
have the form
$$S_{FW} = {1 \over E + m - (\vec k^2 + m^2)^{1/2}} \eqno(4.7)$$
Clearly, this has a pole at
$$E = (\vec k^2 + m^2 )^{1/2} - m \eqno(4.8)$$
with unit residue.

On the other hand, if we write
$$\psi = \pmatrix{\psi_L\cr
\noalign{\vskip 5pt}%
\psi_S\cr} \eqno(4.9)$$
then the Lagrangian of Eq. (4.1$^\prime$) will have the form
$${\cal L} = \psi^\dagger_L  i \partial_0
\psi_L + \psi^\dagger_L i \vec \sigma
\cdot \vec \nabla \psi_S + \psi^\dagger_S i \vec \sigma \cdot
\vec \nabla \psi_L +
\psi^\dagger_S \left( i \partial_0 + 2m \right) \psi_S \eqno(4.10)$$
Since the Lagrangian is at the most quadratic in the fermions, we can
integrate out the ``small" component and obtain
$$\eqalign{Z &= N \int {\cal D} \psi^\dagger_L {\cal D} \psi_L {\cal D}
\psi^\dagger_S {\cal D} \psi_S
\ e^{i \int d^4x
\ {\cal L}}\cr
&= N [ \det
( i \partial_0 + 2m)] \int {\cal D} \psi^\dagger_L {\cal D}
\psi_L\
e^{i \int d^4x  \psi^\dagger_L (i \partial_0 +
( i \partial_0 + 2m)^{-1}  \vec \nabla^2 ) \psi_L}\cr
&= N [ \det (i \partial_0 + 2m)] \ Z_{\rm CONV}\cr}\eqno(4.11)$$
\noindent From the
 structure of the Lagrangian in Eq. (4.11) we note that the
conventional (naive)
propagator for the ``large" component will have the form
$$S_{\rm CONV} = {1 \over E - {\vec k^2 \over E+2m}} \eqno(4.12)$$
We note that the propagator has poles at
$$E = \pm \left( \vec k^2 + m^2 \right)^{1/2} - m\eqno(4.13)$$
The location of the positive energy pole, in which we are interested,
coincides with that of the other approach.  But we note that the residue at
the positive energy pole is ${m + (\vec k^2 + m^2)^{1/2} \over
2(\vec k^2 + m^2)^{1/2}}$.  This suggests that the field $\psi_L$ needs to
be redefined.  Note that if we redefine (compare this with
Eq. (3.16$^{\prime \prime}$))
$$\psi_L \rightarrow \widetilde \psi_L = \left( i \partial_0 + 2m \right)^{-
1/2}  \left( i \partial_0 + m
 + \left( m^2 - \vec \nabla^2 \right)^{1/2}
\right)^{1/2} \psi_L \eqno(4.14)$$
then the generating functional will have the form
$$\eqalign{Z &= N [\det
(i \partial_0 + 2m) ] \int {\cal D} \psi^\dagger_L
{\cal D} \psi_L
e^{i \int d^4x \ \psi^\dagger_L
 (i \partial_0  + (i \partial_0 +2m)^{-1}
\vec \nabla^2 )\psi_L}\cr
&= N [ \det
( i \partial_0 + 2m)]
[\det (i \partial_0 +2m)]^{-1}\cr
&\qquad\qquad \times [\det(i \partial_0 + m
 + (m^2 - \vec \nabla^2)^{1/2})]
\int {\cal D} \widetilde \psi^\dagger_L {\cal D}
\widetilde \psi_L
e^{i \int d^4x\ \widetilde  \psi^\dagger_L (i \partial_0 + m -
(m^2 -
\vec \nabla^2 )^{1/2})\widetilde \psi_L}\cr
&= N [ \det (i \partial_0 + m + (m^2 - \vec
\nabla^2)^{1/2})]\  \widetilde  Z_{\rm CONV}\cr}\eqno(4.15)$$
Comparing Eqs. (4.6) and (4.15), we conclude that
$$Z_{FW} = \widetilde Z_{\rm CONV} \eqno(4.16)$$
Namely, the two approaches lead to equivalent theories but only after
appropriate field redefinition.  For an interacting theory,
Foldy-Wouthuysen transformations do not have a closed form.  Therefore, the
equivalence in the functional approach has to be shown order by order.
An an example, let us consider the Dirac particle interacting with a
constant electric field up to order ${1 \over m^2}$.  The Lagrangian with
the phase redefinition of Eq. (4.1$^\prime$) is given by
$${\cal L} = \overline \psi \left( i \gamma^0 \left( \partial_0 + ie A_0
\right) + i \vec \gamma \cdot \vec \nabla - m \left( 1 -
\gamma^0 \right) \right) \psi \eqno(4.17)$$
In the Foldy-Wouthuysen approach, if we diagonalize the Lagrangian
up  to order
${1 \over m^2}$, it will have the form (FW transformations are unitary and
hence the functional measure does not change.)
$$\eqalign{{\cal L} = &\overline \psi^\prime
\bigg( i \gamma^0 \big( \partial_0 + i
e A_0 \big) - m \big( 1 - \gamma^0 \big)\cr
& + {1 \over
2m}\ \vec \nabla^2 - {ie \over 4m^2}\ \gamma_5
\vec \gamma \cdot \big( \vec E \times \vec \nabla \big)
+ O \big( {1 \over m^3}\big) \bigg) \psi^\prime
\cr} \eqno(4.18)$$
Writing $\psi^\prime$ as in Eq. (4.3), the Lagrangian density has the form
$$\eqalign{{\cal L} &= \psi^\dagger_1 \bigg( i \partial_0 - e A_0
+ {1 \over 2m}\ \vec \nabla^2 + {ie \over 4m^2} \ \vec \sigma \cdot
\big( \vec E \times \vec \nabla \big) + O \big( {1 \over m^3}\big) \bigg)
\psi_1\cr
 &+ \psi^\dagger_2 \bigg( i \partial_0 - e A_0
+ 2m - {1 \over 2m}\  \vec \nabla^2 + {ie \over 4m^2} \ \vec \sigma \cdot
\big( \vec E \times \vec \nabla \big) + O \big( {1 \over m^3}\big) \bigg)
\psi_2\cr}\eqno(4.19)$$
The generating functional, therefore, has the form
$$\eqalign{Z &= N \int {\cal D} \psi^\dagger_1 {\cal D} \psi_1 {\cal D}
\psi^\dagger_2 {\cal D} \psi_2 e^{i \int d^4x\ {\cal L}}\cr
 &= N \bigg[ \det \bigg( i \partial_0 -
e A_0 + 2m - {1 \over 2m}\
 \vec \nabla^2 + {ie \over 4m^2} \ \vec \sigma \cdot
\big( \vec E \times \vec \nabla \big) + O \big( {1 \over m^3}\big) \bigg)
\bigg]\cr
&\qquad \times \int {\cal D} \psi^\dagger_1 {\cal D} \psi_1 e^{i \int d^4x\
\psi_1^\dagger \big(
i \partial_0 - eA_0 + {1 \over 2m}\ \vec \nabla^2 + {ie \over 4m^2}
\vec \sigma \cdot \big(
\vec E \times \vec \nabla \big) + O \big( {1 \over m^3}\big) \big)
\psi_1}\cr
&= N \bigg[ \det \bigg( 2m \big( 1 + {1 \over 2m}\ \big( i \partial_0 - e
 A_0 \big) - {1 \over 4m^2}\ \vec \nabla^2 + O
\big( {1 \over m^3}\big)\big) \bigg) \bigg] \ Z_{FW} \cr}\eqno(4.20)$$

On the other hand, the Lagrangian in terms of ``large" and ``small"
components has the form
$${\cal L} = \psi^\dagger_L \left( i \partial_0 - e A_0 \right) \psi_L
+ \psi^\dagger_L
i \vec \sigma \cdot
\vec \nabla \psi_S +
 \psi^\dagger_S i \vec \sigma \cdot
\vec \nabla \psi_L + \psi^\dagger_S \left(
i \partial_0 - e A_0 + 2m \right) \psi_S
\eqno(4.21)$$
Integrating out the ``small" components we have
$$\eqalign{Z &= N \big[ \det ( i \partial_0 - e A_0 + 2m ) \big]
\int {\cal D} \psi^\dagger_L {\cal D} \psi_L
e^{i \int d^4x\ \psi^\dagger_L \big( i \partial_0 - e
A_0 + \vec \sigma \cdot \vec \nabla
( i \partial_0 - e A_0 + 2 m )^{-1}
\vec \sigma \cdot \vec \nabla \big) \psi_L}\cr
 &= N \bigg[ \det
 \bigg( 2m \big( 1 + {1 \over 2m} \big( i \partial_0
-eA_0 \big) \big) \bigg) \bigg]
\int {\cal D} \psi^\dagger_L {\cal D} \psi_L\cr
&\qquad\qquad \times e^{i \int d^4x\ \psi^\dagger_L \big( i \partial_0 - e
A_0 + {1 \over 2m}\ \vec \nabla^2 - {e \over 4m^2}\
\vec \sigma \cdot \vec \nabla
\vec \sigma \cdot \vec E - {1 \over 4m^2}\ \vec \nabla^2 \big(
i \partial_0 - eA_0 \big) \big) \psi_L}\cr
&= N \bigg[ \det \bigg( 2m \big( 1 + {1 \over 2m}\ \big( i \partial_0 - e A
_0 \big) \big) \bigg) \bigg] \ Z_{\rm CONV} \cr} \eqno(4.22)$$
If we now redefine (as in Eq. (2.14))
$$\psi_L \rightarrow \widetilde \psi_L = \left(
1 - {1 \over 8m^2}\ \vec \nabla^2 \right) \psi_L \eqno(4.23)$$
then the Lagrangian in Eq. (4.22) would take the form
$${\cal L}_{\rm CONV} = \widetilde \psi^\dagger_L \left( i \partial_0 -
e A_0
+ {1 \over 2m}\ \vec \nabla^2 + {ie \over 4m^2}\ \vec \sigma \cdot
\left( \vec E \times \vec \nabla \right) + O \left( {1 \over m^3}\right)
\right) \widetilde \psi_L \eqno(4.24)$$
Furthermore, with the Jacobian arising from the redefinition in Eq. (4.23),
the generating functional would become
$$\eqalign{Z &= N \bigg[\det \bigg( 2m \big( 1 + {1 \over 2m}\ \big( i
\partial_0 - e A_0 \big) \big) \bigg) \bigg]
\bigg[ \det \bigg( 1 - {1 \over 8m^2} \ \vec \nabla^2 \bigg)
\bigg]^2\cr
&\qquad \times
\int {\cal D} \widetilde \psi_L^\dagger {\cal D} \widetilde \psi_L
e^{i \int d^4x\ \widetilde \psi^\dagger_L
\bigg( i \partial_0 - eA_0 + {1 \over 2m}\
\vec \nabla^2 + {ie \over 4m^2}\ \vec \sigma \cdot
\big( \vec E \times \vec \nabla \big)
+ O \big( {1 \over m^3}\big)
\bigg) \widetilde \psi_L}\cr
 &= N \bigg[ \det \bigg( 2m \big( 1 + {1 \over 2m}\ \big( i
\partial_0 - e A_0 \big) - {1 \over 4m^2}\
\vec \nabla^2 + O \big( {1 \over m^3}\big) \big)\bigg)
\bigg]\ \widetilde Z_{\rm CONV} \cr} \eqno(4.25)$$
Comparing Eqs. (4.20) and (4.25), we conclude that
$$Z_{FW} = \widetilde Z_{\rm CONV} \eqno(4.26)$$
This establishes equivalence of the two approaches up to order
${1 \over m^2}$ in the presence of a constant electric field.  Once again,
the significance of the field redefinition cannot be over emphasized.  It is
tedious, but as the discussion of the earlier section shows, the
equivalence can be carried out to any given order  with appropriate field
redefinitions.  The higher order equivalence, however, can only be shown
through a careful iterative procedure.

\medskip

\noindent {\bf V. Conclusion:}

\medskip

In this note we have shown up to order ${1 \over m^5}$ that the two ways
of obtaining an interacting heavy quark effective theory yield the same
result.  From a practical point of view, it is the Foldy-Wouthuysen method
that is simpler for the derivation of
 the effective theory.  However, since it involves
field dependent unitary transformations, unless the fields vanish
sufficiently rapidly, it is conceptually unclear whether the S-matrix
elements will remain the same [17].  The traditional method of eliminating the
``small" components, on the other hand, is quite tricky.  Here the
``large" component fields must be renormalized
 (redefined) and this must be carried out
carefully in an iterative manner.  This is essential since otherwise the
calculations may not represent the true physical effects.  This procedure
of field redefinition is quite tedious but has the virtue that the
existence of the S-matrix elements is never in question in this approach
since the asymptotic form of the wave function
remains unchanged in general.
Thus, in some sense, one can view the two approaches to be complementary
and equivalent.

\noindent {\bf Acknowledgement}

\medskip
I would like to thank Dr. D. Pirjol for raising the question of equivalence
of the two approaches.  I would like to thank him as well as Dr. M. Neubert
 for sending their papers before publication.  It is a great pleasure to
thank Prof. L. L. Foldy and in particular Prof. S. Okubo for many helpful
comments and suggestions.  This work was supported in part by U.S.
Department of Energy Grant no. DE-FG-02-91ER40685.
\vfill\eject
\baselineskip=18pt
\noindent {\bf References:}

\item{1.} W. E. Caswell and G. P. Lepage, Phys. Lett. {\bf B167} (1986)
437; G. P. Lepage and B. A. Thacker, Nucl. Phys. B (Proc. Supp.) {\bf 4}
(1988) 199.

\item{2.} N. Isgur and M. B. Wise, Phys. Lett. {\bf B232} (1989) 113;
{\it ibid} {\bf B237} (1990) 527.

\item{3.} E. Eichten and B. Hill, Phys. Lett. {\bf B234} (1990) 511; M. E.
Luke, Phys. Lett. {\bf B252} (1990) 447; A. F. Falk and B. Grinstein, Phys.
Lett. {\bf B247} (1990) 406; {\it ibid} {\bf B249} (1990) 314; A. F. Falk,
B. Grinstein and M. E. Luke, Nucl. Phys. {\bf B357} (1991) 185; A. F. Falk,
H. Georgi and B. Grinstein, Nucl. Phys. {\bf B343} (1990) 1; H. Georgi,
Nucl. Phys. {\bf B348} (1991) 293; H. Georgi, B. Grinstein and M. B. Wise,
Phys. Lett. {\bf B252} (1990) 456.

\item{4.} H. Georgi, Phys. Lett. {\bf B240} (1990) 447.

\item{5.} B. Grinstein, HUTP-91/A028 (1991); HUTP-91/A040 (1991) and SSCL-
17 (1991).

\item{6.} For a more recent review see M. Neubert, SLAC-PUB-6263 (1993)
to appear in Physics Reports.

\item{7.} L. L. Foldy and S. A. Wouthuysen, Phys. Rev. {\bf 78} (1950) 29.

\item{  } S. Tani, Prog. Theor. Phys. {\bf 6} (1951) 267.

\item{8.} J. G. K\"orner and G. Thompson, Phys. Lett. {\bf B264} (1991)
185.

\item{9.} T. Mannel, W. Roberts and Z. Ryzak, Nucl. Phys. {\bf B368} (1992)
204.
\item{  } J. Soto and R. Tzani, Phys. Lett. {\bf B297}
(1992) 358.

\item{10.} S. Balk, J. G. K\"orner and D. Pirjol,
 MZ-TH/93-13 (1993).

\item{11.} A. F. Falk and M. Neubert, Phys. Rev. {\bf D47} (1993) 2965.
Also see ref. 6.

\item{12.} W. Pauli, {\it Handbuch der Physik}, Vol. V/1,  (Heidelberg,
1958).

\item{13.} K. Sawada, unpublished; N. Fukuda, K. Sawada and M. Taketani,
Prog. Theor. Phys. {\bf 12} (1954) 156.

\item{14.} S. Okubo, Prog. Theor. Phys. {\bf 12} (1954) 102, 603.

\item{15.} M. Sugawara and S. Okubo, Phys. Rev. {\bf 117} (1960) 605, 611.

\item{16.} See for example refs. 7, 8 and A. Das and V. S. Mathur, UR-1320
(1993).

\item{17.} S. Kamefuchi, L. O'Raifeartaigh and A. Salam, Nucl. Phys.
{\bf 28} (1961) 529.

\end